\documentclass[twocolumn,showpacs,preprintnumbers,amsmath,amssymb]{revtex4}

\usepackage{graphicx}
\usepackage{dcolumn}
\usepackage{bm}

\begin{document}

\title{First results on the interactions of relativistic $^9$C nuclei in nuclear track emulsion}
\author{D.~O.~Krivenkov}
   \affiliation{Joint Insitute for Nuclear Research, Dubna, Russia} 
\author{D.~A.~Artemenkov}
   \affiliation{Joint Insitute for Nuclear Research, Dubna, Russia} 
 \author{V.~Bradnova}
   \affiliation{Joint Insitute for Nuclear Research, Dubna, Russia} 
\author{M.~Haiduc}
   \affiliation{Institute of Space Sciences, Magurele, Romania}
\author{S.~P.~Kharlamov}
   \affiliation{Lebedev Institute of Physics, Russian Academy of Sciences, Moscow, Russia}
\author{V.~N.~Kondratieva}
   \affiliation{Joint Insitute for Nuclear Research, Dubna, Russia}   
\author{A.~I.~Malakhov}
   \affiliation{Joint Insitute for Nuclear Research, Dubna, Russia} 
\author{A.~A.~Moiseenko}
   \affiliation{Yerevan Physics Institute, Yerevan, Armenia}
\author{G.~I.~Orlova}
   \affiliation{Lebedev Institute of Physics, Russian Academy of Sciences, Moscow, Russia} 
\author{N.~G.~Peresadko}
   \affiliation{Lebedev Institute of Physics, Russian Academy of Sciences, Moscow, Russia}   
\author{N.~G.~Polukhina}
   \affiliation{Lebedev Institute of Physics, Russian Academy of Sciences, Moscow, Russia}    
\author{P.~A.~Rukoyatkin}
   \affiliation{Joint Insitute for Nuclear Research, Dubna, Russia}  
\author{V.~V.~Rusakova}
   \affiliation{Joint Insitute for Nuclear Research, Dubna, Russia} 
\author{V.~R.~Sarkisyan}
   \affiliation{Yerevan Physics Institute, Yerevan, Armenia} 
\author{R.~Stanoeva}
   \affiliation{Joint Insitute for Nuclear Research, Dubna, Russia} 
\author{T.~V.~Shchedrina}
   \affiliation{Joint Insitute for Nuclear Research, Dubna, Russia} 
\author{S.~Vok\'al}
   \affiliation{P. J. \u Saf\u arik University, Ko\u sice, Slovak Republic}  
\author{P.~I.~Zarubin}
     \email{zarubin@lhe.jinr.ru}    
     \homepage{http://becquerel.lhe.jinr.ru}
   \affiliation{Joint Insitute for Nuclear Research, Dubna, Russia} 
 \author{I.~G.~Zarubina}
   \affiliation{Joint Insitute for Nuclear Research, Dubna, Russia}

\date{\today}

\begin{abstract}
\indent First results of the exposure of nuclear track emulsions in a secondary beam enriched by $^9$C nuclei at energy of 1.2~A~GeV are described. The presented statistics corresponds to the most peripheral $^9$C interactions. For the first time a dissociation $^9$C $\rightarrow$3$^3$He not accompanied by target fragments and mesons is identified.\par
\end{abstract}
\maketitle

\section{\label{sec:level1}Introduction}
\indent The most peripheral processes of the fragmentation of relativistic nuclei on heavy nuclei of the emulsion composition (i. e., Ag and Br) proceed without production of target fragments and mesons. They are called \lq\lq white\rq\rq stars aptly reflecting the images of events \cite {Baroni90,Baroni92}. The fraction of such events that are induced by electromagnetic diffraction and nuclear interactions is a few percent of inelastic interactions. The statistics of various configurations of relativistic fragments reflects the cluster features of light nuclei due to minimal transferred excitation \cite{Andreeva05,Belaga95,Adamovich99,Artemenkov07,Shchedrina07,Peresadko07,Stanoeva07,ArtemenkovAIP07}. The use of emulsion provides a complete monitoring of relativistic fragments with an excellent angular resolution. This approach to the study of the nucleon clustering is used by the BECQUEREL collaboration \cite{Web} for the study of light nuclei at the proton drip line. Exploration of the dissociation of lighter nuclei $^7$Be \cite{Peresadko07} and $^8$B \cite{Stanoeva07} formed the basis for the progress in the study of the next nucleus - $^9$C. One can expect that in the peripheral $^9$C dissociation the picture hitherto obtained for $^8$B and $^7$Be with the addition of one or two protons, respectively, should be reproduced.\par
\indent The $^3$He based clustering plays an equally important role in these nuclei as the $\alpha$-particle one does. For the $^9$C nucleus a cluster excitation 3$^3$He with a relatively low threshold (around 16~MeV) becomes available. In this case, a rearrangement of a neutron from the $\alpha$-particle cluster into the emerging $^3$He nucleus should occur. The search for the dissociation $^9$C$\rightarrow$3$^3$He without accompanying fragments of the target and mesons, i. e. \lq\lq white\rq\rq stars, becomes the main task of this study. In principle, this bright channel could be identified by a trident of doubly charged fragments. But the real situation with emulsion exposure in the secondary beams of relativistic radioactive nuclei is more complicated. There should be a detailed analysis of events of the \lq\lq white\rq\rq star type as the most clearly interpreted interactions for a reliable determination of the used beam composition. In what follows, first results on $^9$C identified interactions are described.\par

\section{\label{sec:level2}Experiment}
\begin{figure*}
\begin{center}
\includegraphics[width=135mm]{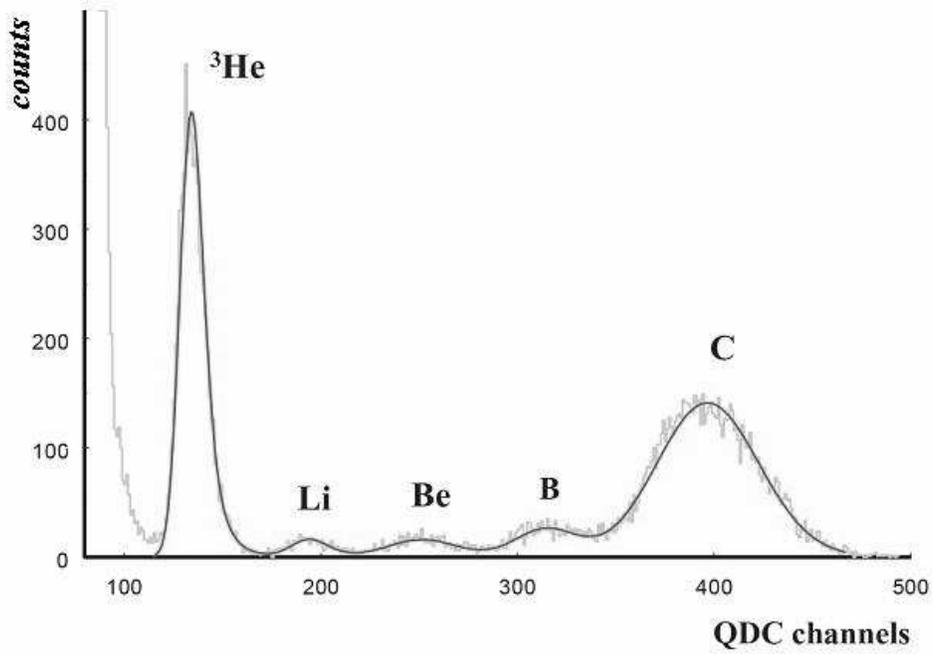}
\caption{\label{fig:1}Charge spectrum of nuclei produced in the fragmentation of $^{12}$C$\rightarrow ^9$C at secondary beam tuning Z$_{pr}$/A$_{pr}$ = 2/3}
\end{center}
\end{figure*}

\indent The fragmentation of 1.2~A~GeV $^{12}$C nuclei, accelerated at the JINR Nuclotron, was used to form a secondary beam with low magnetic rigidity for the best selection of $^9$C nuclei \cite{Rukoyatkin06}. The momentum acceptance of the separating channel was about 3\%. The amplitude spectrum of a scintillation monitor of the secondary beam is presented Fig.~\ref{fig:1}. It shows that the major contribution comes from C nuclei. The main background was an admixture of nuclei $^3$He, which have the same ratio of the charge Z$_{pr}$ to the atomic mass A$_{pr}$, as $^9$C ones have. $^4$He nuclei could not penetrate into the channel because of a much greater magnitude of this ratio. A small admixture of fragments $^7$Be and $^8$B with slightly higher magnetic rigidity than that of $^9$C entered the beam. These spectrum features indicate the correctness of the channel tuning.\par
\indent The exposed stack consisted of 19 layers of BR-2 nuclear track emulsion with a relativistic sensitivity. The layer thickness and dimensions were 0.5 $mu$m and 10$\times$20 cm$^2$. Stack exposure was performed in a beam directed in parallel to the plane of the stack along the long side. The presented analysis is based on a complete scanning of 13 layers along the primary tracks with charges visually assessed as Z$_{pr}>$ 2. $^3$He nuclei were rejected at the primary stage of selection. The ratio of intensities Z$_{pr}>$ 2 and Z$_{pr}$ = 2 was about 1:10. The presence of particles Z$_{pr}$ = 1 in the ratio Z$_{pr}>$ 2 1:1 was also detected. The Z$_{fr}$ = 1 fragments were separated visually from the Z$_{fr}$ = 2 fragments, because their ionization was four times smaller. Over the viewed length of 167.1~m traces one found 1217 interactions mostly produced by C nuclei. Thus, it was obtained that the mean free path was equal to $\lambda_{C}$ = 13.7$\pm$0.4~cm. This value corresponds to the estimate based on the data for the neighboring cluster nuclei.\par
\begin{figure*}
\begin{center}
\includegraphics[width=135mm]{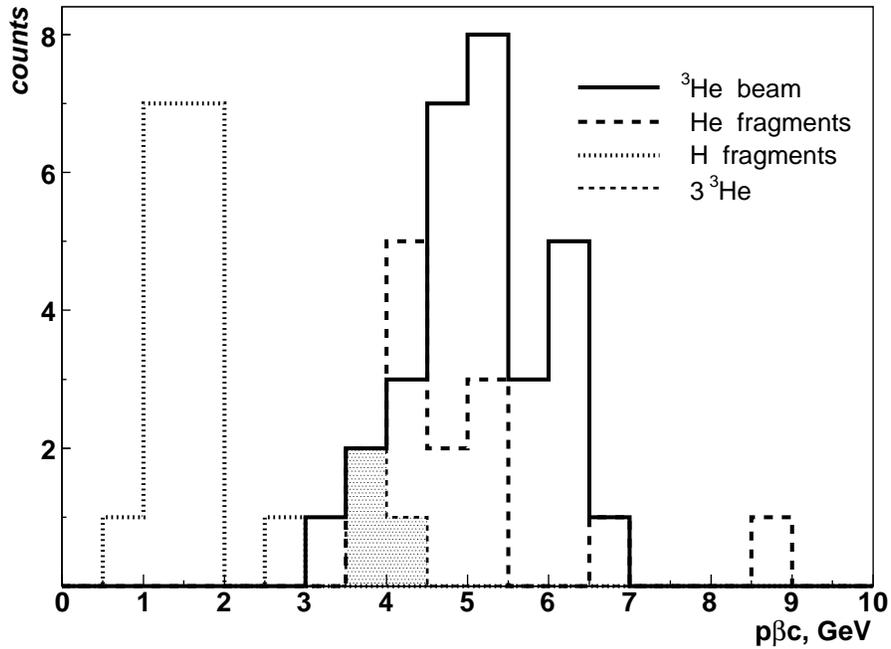} 
\caption{\label{fig:2}Distribution of the measurements p$\beta$c for beam $^3$He nuclei (30 tracks, solid histogram), singly charged fragments of the \lq\lq white\rq\rq stars $\Sigma$Z$_{fr}$ = 5 + 1 and 4 + 1 + 1 (15 tracks, dot histogram), doubly charged fragments of the \lq\lq white\rq\rq stars 3He (14 traces, dotted histogram) and from the 3$^3$He event (shaded histogram)}
\end{center}
\end{figure*}
\indent The relativistic fragments H and He can be identified in the cases of small variations in their values p$\beta$c, derived from measurements of multiple scattering, where p is the full momentum, and $\beta$ the speed. It is assumed that the projectile fragments conserve their momentum per nucleon, i. e., p$\beta$c$\approx$A$_{fr}$p$_{0}\beta_{0}$c, where A$_{fr}$ the fragment atomic number. To achieve the required precision one needs to measure the deflection of the track coordinates in more than 100 points. Despite the workload, this method provides a unique completeness of the information on the composition of the systems of few lightest nuclei.\par
\indent The presence of $^3$He nuclei in the beam composition was found to be helpful to calibrate the identification conditions for secondary fragments. The distribution of the p$\beta$c measurements for 30 nuclei $^3$He from the beam is presented in Fig.~\ref{fig:2} (solid histogram). The average value of the distribution is $<p\beta c>$= 5.1$\pm$0.2~GeV in the mean scattering $\sigma$= 0.8~GeV. The absolute value is somewhat different from the expected value of 5.4~GeV for $^3$He nuclei (for $^4$He - 7.2~GeV) and is defined by the traditionally used constants. The $\sigma$ value can be assumed to be satisfactory for the separation of isotopes $^3$He and $^4$He, and especially their systems.\par
\begin{figure*}
\begin{center}
\includegraphics[width=135mm]{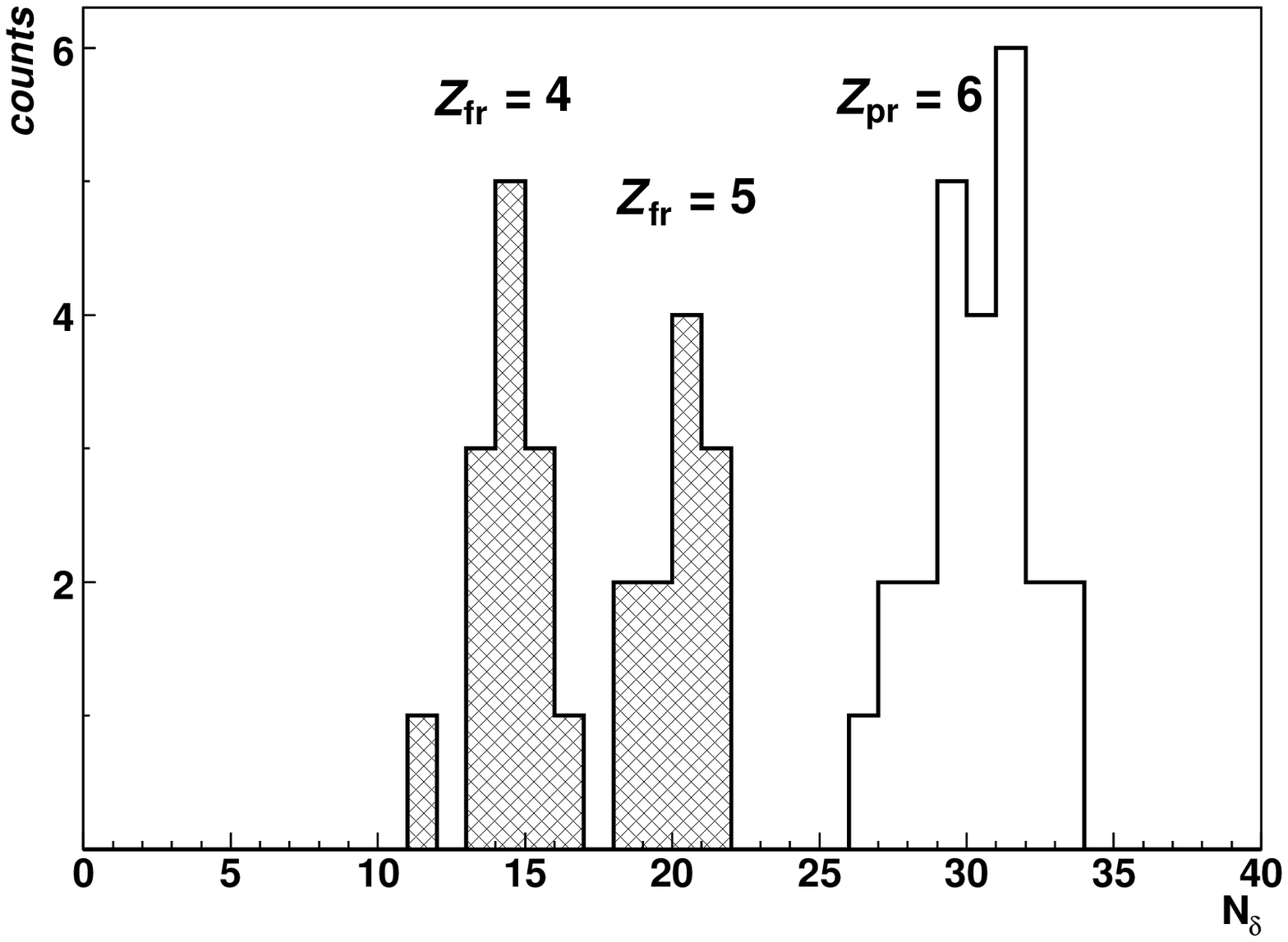}
\caption{\label{fig:3}Distributions by the mean number of $\delta$-electrons per 1 mm length for beam particles (solid histogram) and relativistic fragments with charges Z$_{fr}>$ 2 (shaded histogram) from \lq\lq white\rq\rq stars $\Sigma$Z$_{fr}$~= 5 + 1 and 4 + 1 + 1}
\end{center}
\end{figure*}
\indent The contribution of the C, Be and B isotopes was separated via the charge configurations of secondary fragments $\Sigma$Z$_{fr}$ in \lq\lq white\rq\rq stars and subsequent measurements of the primary charges Z$_{pr}$. The charges of the projectile nuclei and fragments Z$_{fr}>$ 2 were determined by counting of $\delta$-electrons on tracks. The measurement of the charges of the primary nuclei and fragments of the events $\Sigma$Z$_{fr}$ = 5 + 1 and 4 + 1 + 1, presented in Fig.~\ref{fig:3}, allows one to conclude that all the events are originated from nuclei Z$_{pr}$ = 6. There is an expected shift in the distribution for interaction fragments.\par

\section{\label{sec:level3}Fragment charged configurations}
\indent The distribution of 123 \lq\lq white\rq\rq stars N$_{ws}$ in the charge configurations $\Sigma$Z$_{fr}$ is presented in Table~\ref{tab:1}. Events with fragments Z$_{fr}$ = 5 and 4 and identified charges Z$_{pr}$ = 6, accompanied by protons, are interpreted as channels $^9$C$\rightarrow ^8$B + p and $^7$Be + 2p, due to the absence of stable isotopes $^9$B and $^8$Be. These two channels are relative to the most low-threshold dissociation of the nucleus $^9$C and constitute about 30\% of the events of $\Sigma$Z$_{fr}$ = 6. The ratio of $^8$B \lq\lq white\rq\rq stars with heavy fragments ($^8$B$\rightarrow ^7$Be + p) and stars containing only H and He are shown to be approximately equal \cite{Stanoeva07}. Therefore, one can expect that the statistics of the Table 1 contains a large fraction of events produced exactly by the $^9$C nuclei.\par
\begin{figure*}
\begin{center}
\includegraphics[width=135mm]{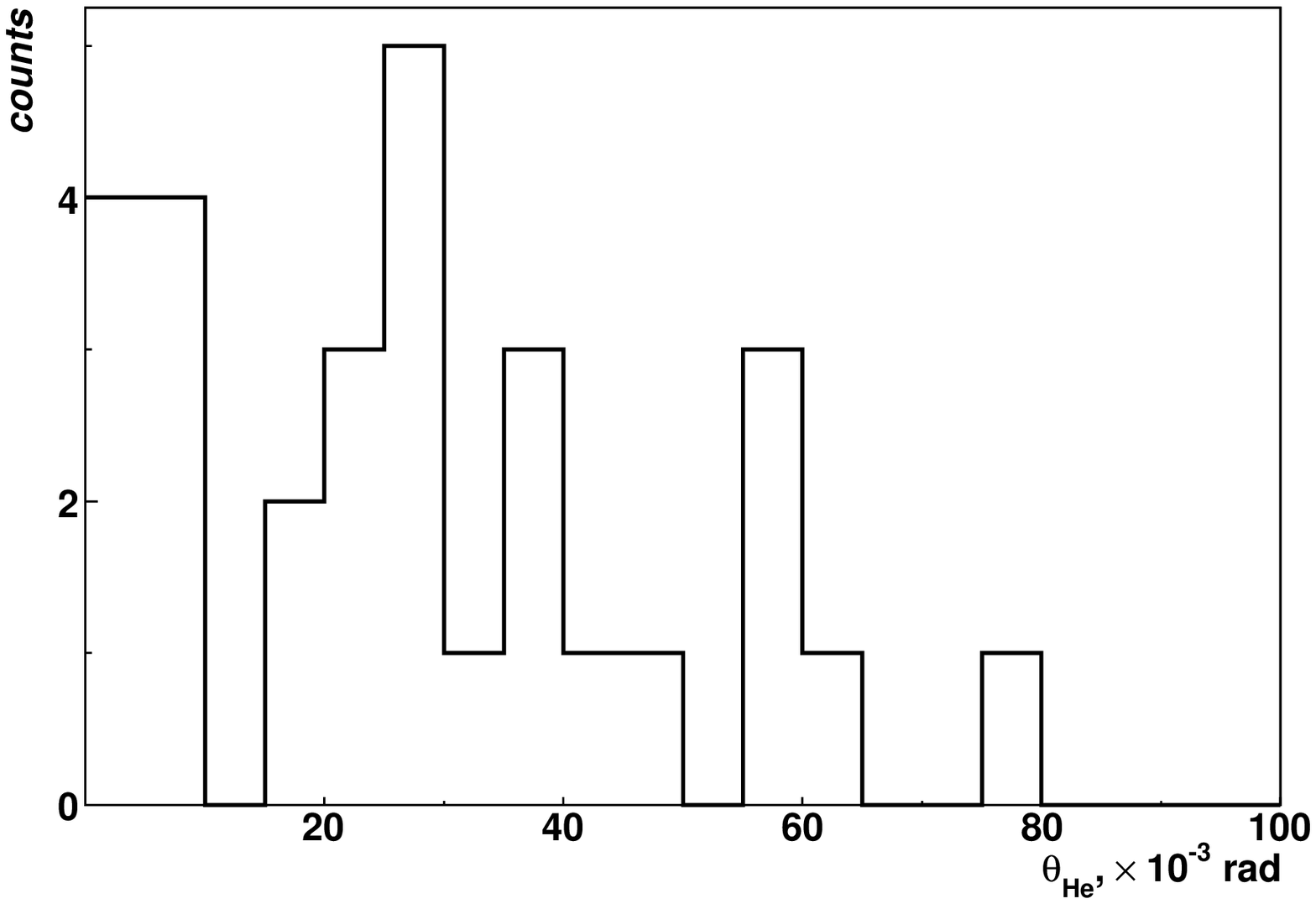}
\caption{\label{fig:4}Distribution by polar angles $\theta$ for doubly charged fragments in the \lq\lq white\rq\rq stars C$\rightarrow$3He}
\end{center}
\end{figure*}
\indent The result of identification of fragments Z$_{fr}$ = 1 in this group of events is presented in Fig.~\ref{fig:4} (dotted histogram). The distribution has $<p\beta c>$ = 1.5$\pm$0.1~GeV and $\sigma$ = 0.4~GeV. In fact, the identification in these cases is not necessary because of limited options, and protons can serve as a calibration mechanism in more complicated cases.\par
\indent Table~\ref{tab:1} allows one to derive useful indirect conclusions about the composition of the beam. For example, there is only one event $\Sigma$Z$_{fr}$ = 4 + 2, which might arise from the dissociation $^{11}$C$\rightarrow ^7$Be + $^4$He having the lowest threshold for the isotope $^{11}$C. Thus, a possible presence of $^{11}$C in the secondary beam is negligible. Six \lq\lq white\rq\rq stars with the total charge $\Sigma$Z$_{fr}$ = 7 are associated with $^{12}$N nuclei, captured in the beam. $^{12}$N nuclei were produced in charge exchange processes $^{12}$C$\rightarrow ^{12}$N in the producing target.\par
\indent The distribution of \lq\lq white\rq\rq stars produced by $^7$Be, $^8$B and C nuclei of the charge configurations $\Sigma$Z$_{fr}$, which consist only of the nuclei H and He, is presented in the Table~\ref{tab:2}. One excluded from the sum $\Sigma$Z$_{fr}$ one H nucleus for the $^8$B cases and 2H – for the C cases. There are similar fractions of the channels 2He and He + 2H, which correspond to the appearance of $^7$Be as a core of $^9$C.\par
\indent Besides, it is possible to note the production of five \lq\lq white\rq\rq stars C$\rightarrow$6H (Table~\ref{tab:2}). Events of this type in the cases of the $^{12}$C and $^{11}$C isotopes, requiring a simultaneous collapse of three He clusters, could not practically proceed without target fragments due to a very high threshold. Being related to an extremely high threshold, they could not proceed without the production of target fragments. In contrast, similar processes related to the breakups of cluster He pairs, were observed for the \lq\lq white\rq\rq stars $^7$Be$\rightarrow$4H \cite{Peresadko07} and $^8$B$\rightarrow$5H \cite{Stanoeva07}. These events require a full identification of the He and H isotopes and a kinematical analysis to be performed in future. It is quite possible that one will find some cases which would correspond to the cross-border stability in the direction of nuclear resonance states $^9$C$\rightarrow ^8$C.\par
\indent The question about the contribution of $^{10}$C nuclei to the statistics of the Table 1 which would be characterized by configurations consisting only of the He and H isotopes, remains open. A detailed identification and an angular analysis will be crucial in this regard.\par
\section{\label{sec:level4} Search for 3$^3$He events}
\begin{figure*}
\begin{center}
\includegraphics[width=135mm]{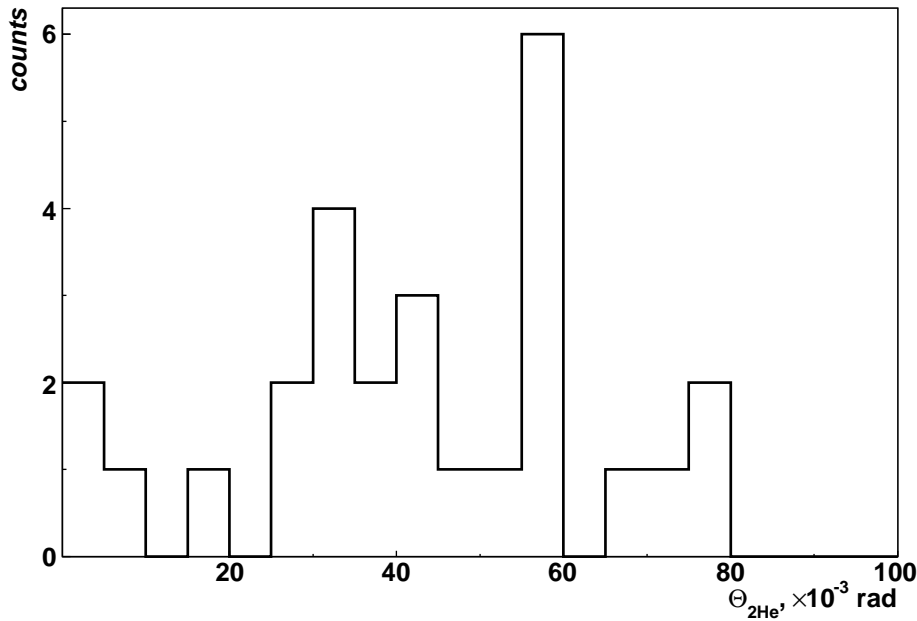}
\caption{\label{fig:5}Distribution by opening angles $\Theta_{2He}$ between fragments in the \lq\lq white\rq\rq stars C$\rightarrow$3He.}
\end{center}
\end{figure*}

\indent
Table 1 shows the production of 13 \lq\lq white\rq\rq stars with the 3He configuration. The distribution by the polar angles $\theta$ for these fragments, which illustrates the degree of collimation, is presented in Fig.~\ref{fig:4}. Angular measurements allow one to derive the opening angle distribution $\Theta_{2He}$ for fragment pairs (Fig.~\ref{fig:5}). Four narrow He pairs with $\Theta_{2He}<10^{-2}$ rad are clearly seen due to an excellent spatial resolution. The obtained statistics permits to begin the search for the channel $^9$C$\rightarrow$3$^3$He as the most interesting challenge. The $^{10}$C admixture could also lead to the events of a deep nucleonic regrouping $^{10}$C$\rightarrow$2$^3$He + $^4$He.
\par
\begin{table}
\caption{\label{tab:1}Distribution of \lq\lq white\rq\rq stars N$_{ws}$ in charge configurations $\Sigma$Z$_{fr}$}
\label{tab:1}       
\begin{tabular}{l|cccccc|c}
\hline\noalign{\smallskip}
$\Sigma$Z$_{fr}$&  &  &  & Z$_{fr}$ &  &  & N$_{ws}$ \\
 & 6 & 5 & 4 & 3 & 2 & 1 & \\
\noalign{\smallskip}\hline\noalign{\smallskip}
7 & - & - & - & - & 1 & 5 & 2 \\
7 & - & - & - & - & 2 & 3 & 4 \\
6(Z$_{pr}$=6) & - & 1 & - & - & - & 1 & 11 \\
6(Z$_{pr}$=6) & - & - & 1 & - & - & 2 & 13 \\
6 & - & - & - & - & 3 & - & 13 \\
6 & - & - & 1 & - & 1 & - & 1 \\
6 & - & - & - & 1 & 1 & 1 & 2 \\
6 & - & - & - & 1 & - & 3 & 2 \\
6 & - & - & - & - & 1 & 4 & 22 \\
6 & - & - & - & - & 2 & 2 & 21 \\
6 & - & - & - & - & - & 6 & 5 \\
5 & - & - & 1 & - & - & 1 & 2 \\
5 & - & - & - & 1 & - & 2 & 2 \\
5 & - & - & - & - & 1 & 3 & 13 \\
5 & - & - & - & - & 2 & 1 & 7 \\
4 & - & - & - & - & 1 & 2 & 4 \\
\noalign{\smallskip}\hline
\end{tabular}
\end{table}
\indent The method of the multiple Coulomb scattering was used to determine p$\beta$c values of He fragments. Such measurements were implemented only for 14 tracks because of the characteristics of the stack used and because of the fragment angular dispersions (dashed histogram in Fig.~\ref{fig:2}). The average value $<p\beta c>$ was found to be equal to 4.6$\pm$0.2~GeV, with $\sigma$= 0.6~GeV. The fraction of the fragments that could be defined as $^4$He nuclei is insignificant as compared with $^3$He. A systematic decrease in the value $<p\beta c>$ with the respect to the beam calibration is due to energy losses in interactions.\par
\begin{table}
\caption{\label{tab:2} Distribution of the numbers and fractions of \lq\lq white\rq\rq stars of $^7$Be, $^8$B and C nuclei, over H and He configurations}
\begin{tabular}{lcccccc}
\hline\noalign{\smallskip}
Channel & $^7$Be & Fraction, \% & $^8$B(+H) & Fraction, \% & $^9$C(+2H) & Fraction, \% \\
\noalign{\smallskip}\hline\noalign{\smallskip}
2He & 41 & 43 & 12 & 40 & 21 & 42 \\
He+2H & 42 & 45 & 14 & 47 & 22 & 44 \\
4H & 2 & 2 & 4 & 13 & 5 & 10 \\
Li+H & 9 & 10 & 0 & 0 & 2 & 4 \\
\noalign{\smallskip}\hline
\end{tabular}
\end{table}

\indent The procedure for determining the p$\beta$c values of all three fragments became possible only in a single event C$\rightarrow$3He (shaded histogram in Fig.~\ref{fig:2}). The values allow one to interpret the event as a triple production of $^3$He nuclei with a total value p$\beta$c = 12$\pm$1~GeV. The interpretation of events as $^{10}$C$\rightarrow$3$^3$He + n is unlikely, because in this case one would need to modify not one but a pair of clusters $^4$He by overcoming a threshold of at least 37~MeV in a peripheral interaction without target fragment production.\par
\begin{figure*}
\begin{center}
\includegraphics[width=150mm]{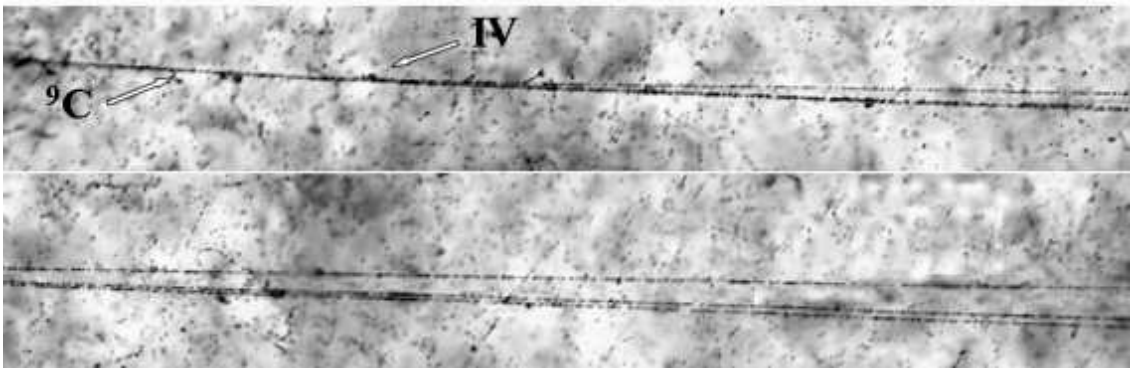}
\caption{\label{fig:6}Microphotography of \lq\lq white\rq\rq star $^9$C$\rightarrow$3$^3$He at 1.2~A~GeV. The upper photo shows the dissociation vertex (indicated as {\bf IV}) and fragments in a narrow cone. Three tracks of relativistic He fragments can clearly be seen in the bottom photo.}
\end{center}
\end{figure*}

\indent This event is a first identified candidate for $^9$C$\rightarrow$3$^3$He. Its mosaic microphotography is presented in Figure~\ref{fig:6}. The values of the 3He transverse momenta P$_{t}$, obtained by p$\beta$c and angular measurements are equal to 318$\pm$53~MeV/c, 128$\pm$20~MeV/c and 110$\pm$14~MeV/c. Then the total momentum $\Sigma$Pt transferred to the system 3$^3$He is equal to 551 ± 60 MeV/c. By introducing the $\Sigma$P$_{t}$ correction, one can get noticeably lower values of P$_{t}^{*}$ in the 3$^3$He center of mass - 138$\pm$23~MeV/c, 66$\pm$10~MeV/c, and 74$\pm$10~MeV/c.\par
\indent The excitation energy estimated from the difference of the invariant mass M$_{eff}$ of the 3$^3$He system and the fragment mass is equal E$_{3He}$ = 11.9$\pm$1.4~MeV. The peculiarity of this event is the presence of a 2$^3$He narrow pair of an energy of E$_{2HE}$ = 46$\pm$8~keV. Such a low E$_{2He}$ value of the relativistic 2$^3$He pair is close to the decay energy from the ground state of $^8$Be$\rightarrow$2$^4$He. In the study of $^7$Be$\rightarrow ^3$He + $^4$He \cite{Peresadko07} so narrow relativistic pairs 2He were not found. The discussed \lq\lq white\rq\rq star is one of the four 3He events with such narrow pairs (Fig.~\ref{fig:5}). This observation can motive the study of an intriguing opportunity of the existence of a 2$^3$He narrow resonant state near the threshold.\par
	
\section{\label{sec:level5}Conclusions}
\indent For the first time nuclear track emulsion is exposed to relativistic $^9$C nuclei. The picture of the charge configurations of relativistic fragments is obtained for the most peripheral interactions. A dissociation event $^9$C$\rightarrow$3$^3$He accompanied by neither target fragments of the nucleus target nor charged mesons is identified. This paper provides a framework for further accumulation of statistics and for a detailed analysis of $^9$C interactions. \par
\begin{acknowledgments}
indent We are very grateful to the JINR Nuclotron staff for the enthusiastic work and friendly cooperation in this experiment. The authors express their particular gratitude to the JINR staff member A. M. Sosulnikova for a large and careful work on visual scanning of emulsions. This work was supported by grants -- 96-1596423, 02-02-164-12a, 03-02-16134, 03-02-17079, 04-02-17151 and 04-02-16593 of Russian fund for basic researches, grants VEGA  \#1/2007/05 and \#1/0080/08 of Agency Science of Ministry of Education of the Slovak Republic and the Slovak Academy of Sciences, as well as grants of plenipotentiaries of Bulgaria, Slovakia, the Czech Republic and Romania in 2002-8. \par
\end{acknowledgments}

\end{document}